\documentclass[sigconf]{acmart}
\usepackage{balance}
\usepackage{flushend}
\usepackage{hyperref}
\usepackage[utf8]{inputenc}
\usepackage[absolute,overlay]{textpos}
\AtBeginDocument{\providecommand\BibTeX{{\normalfont B\kern-0.5em{\scshape i\kern-0.25em b}\kern-0.8em\TeX}}}

\usepackage{wrapfig}

\usepackage{enumitem}  

\usepackage{ulem}
\usepackage{etoolbox} \newtoggle{comments} \toggletrue{comments} \iftoggle{comments}{
    \newcommand{\notejs}[1]
    	{{\color{brown}[{\bf Jat:} #1]}}
    \newcommand{\noteml}[1]
    	{{\color{blue}[{\bf Michelle:} #1]}}
    \newcommand{\notejc}[1]
    	{{\color{cyan}[{\bf Jen:} #1]}}
    \newcommand{\todo}[1]
    	{{\color{red}[{\bf Todo:} #1]}}	
	    	
    \newcommand{\cut}[1]
    	{{\color{red}\sout{#1}}}

}{
    \newcommand{\notejs}[1]{}
    \newcommand{\noteml}[1]{}
    \newcommand{\notejc}[1]{}
    \newcommand{\todo}[1]{}
    \newcommand{\cut}[1]{}

}

\usepackage{cancel}

\usepackage[compact]{titlesec}  \usepackage{titletoc}
\titleformat{\section} [hang]                 {\Large\bfseries} {\thesection}       {1em}                  {\MakeUppercase}       \titlespacing{\section}{0pt}                  {5pt plus 1pt minus 3pt} {4pt plus 2pt minus 3pt}  \titlespacing{\subsection}{0pt}                  {4pt plus 0pt minus 3pt} {4pt plus 0pt minus 3pt}  \titlespacing{\subsubsection}{0pt}                  {4pt plus 0pt minus 3pt} {4pt plus 0pt minus 3pt}  

\begin{document}


\settopmatter{printacmref=false} 
\renewcommand\footnotetextcopyrightpermission[1]{} 
\pagestyle{plain} 

\title[Reviewable Automated Decision-Making: A Framework for Accountable Algorithmic Systems]{Reviewable Automated Decision-Making: \\A Framework for Accountable Algorithmic Systems}

\author{Jennifer Cobbe,  Michelle Seng Ah Lee, Jatinder Singh}
\affiliation{\institution{Compliant and Accountable Systems Research Group,\\ University of Cambridge, UK}}
  \email{firstname.lastname@cst.cam.ac.uk}

\begin{abstract}
This paper introduces \textbf{reviewability} as a framework for improving the accountability of {automated and algorithmic decision-making} (ADM) involving  machine learning. We draw on an understanding of ADM as a socio-technical process involving both human and technical elements, beginning before a decision is made and extending beyond the decision itself. While explanations and other model-centric mechanisms may assist some accountability concerns, they often provide insufficient information of these broader ADM processes for regulatory oversight and assessments of legal compliance. Reviewability involves breaking down the ADM process into technical and organisational elements to provide a \textit{systematic framework} for determining the contextually appropriate record-keeping mechanisms to facilitate meaningful review -- both of individual decisions and of the process as a whole. We argue that a reviewability framework, drawing on administrative law's approach to reviewing human decision-making, offers a practical way forward towards more a more holistic and legally-relevant form of accountability for ADM.
\end{abstract}


\maketitle
\begin{textblock*}{17cm}(2.25cm,25.6cm) 
\textsf{\footnotesize{\textcolor{red}{Cite as: Jennifer Cobbe, Michelle Seng Ah Lee, Jatinder Singh. 2021. Reviewable Automated Decision-Making: A Framework for Accountable Algorithmic Systems. In \textit{ACM Conference on Fairness, Accountability, and Transparency (FAccT `21), March 1--10, 2021, Virtual Event, Canada.} ACM, New York, NY, USA.  \url{https://doi.org/10.1145/3442188.3445921}}}}
\end{textblock*}


\section{Introduction}
\label{sec:introduction}

Recent years have seen growing calls from governments, regulators, and civil society for \textit{automated and algorithmic decision-making (ADM)} to be more transparent and accountable (e.g. ~\cite{commons2018, eurocom2020, adalovelace2020, koene2019, ico2019, demos2020}). Many of these have related not only to information about the algorithms themselves, but also about their development, deployment, and use. Understanding how ADM is commissioned, developed, and operated will be increasingly important as it plays a greater role in society, used by governments, companies, and other organisations alike. Indeed, ADM is already used to make decisions about welfare and housing~\cite{alston2019}, access to credit and financial services~\cite{khandani2010consumer}, and immigration and employment~\cite{kuziemski2020ai}, and is likely to become increasingly widespread. Information on ADM's development and use is crucial for assessing whether decisions have been made lawfully, for ensuring that systems operate properly, and for informing those seeking to protect people from arbitrary or capricious interventions in their lives. 

A particular concern is ADM involving machine learning (ML). Meaningful accountability can be difficult with ML; understanding how complex ML systems work can be challenging, and commercial considerations can conceal broader organisational processes~\cite{burrell2016}. To address this, much research has focused on mechanisms to make models and other technical aspects of ADM interpretable by or explainable to humans in some way (e.g. see~\cite{guidotti2018, arrieta2020}). However, information about models and their workings may not be suitable or relevant for various kinds of accountability, and broader technical and organisational factors are generally under-considered.

This paper introduces the concept of \textbf{reviewability} as an approach to improving the accountability of ADM involving ML (though it is relevant for algorithmic systems in general). This draws on our previous work on applying English administrative law--governing public sector decision-making---to ADM~\cite{anon2019, anon2020}. Administrative law has developed iteratively over decades through judicial review by senior courts,  applying to even the most consequential decisions of life and death. Administrative law as a framework and judicial review as an oversight mechanism understand decision-making as a process beginning before decisions are made and extending to their consequences and effects. Our previous work identified points in the ADM process where administrative law's principles and requirements could be applied, enabling effective judicial review of that process as a whole. As a common law jurisdiction with considerable influence 
on others and which reflects legal frameworks across democratic countries~\cite{venicecom}, and drawing on our previous work, we develop reviewability from English administrative law. As such, references to `administrative law' herein mean English administrative law, unless otherwise stated. 

Others have also applied concepts from administrative law (English or otherwise) to ADM in various ways~\cite{mulligan2019, coglianese2017, citron2014, oswald2018, tomlinson2019, zalnieriute2019}. However, these have generally not viewed ADM as a broad process, or,
as with our previous work, sought to apply legal standards to public sector ADM. We believe that administrative law's understanding of decision-making as a process is also highly relevant to accountability of ADM and algorithmic systems more generally -- in private and public sectors. We do not argue that public sector \textit{standards} for decision-making should apply to the private sector, but that a similarly broad \textit{approach} applied to algorithmic systems---considering the whole decision-making process to identify points of review and potential intervention---would benefit accountability of ADM in any sector. We thus build on our work in this area to move beyond a narrower public sector focus and develop a systematic, holistic framework for reviewable ADM as a process applicable across sectors.

\subsection{A systematic framework for holistic review}

We set out a holistic understanding of ADM as a broad socio-technical process, involving both human and technical elements, beginning with the conception of the system and extending through to use, consequences, and investigation. As ADM often involves multiple people working at different stages of the process (together or separately), we consider these human elements as being \textit{organisational} in nature. That is, the human elements of ADM---actions and decisions of people---generally take place in the context of, are informed by, and constitute organisational systems and processes. We note that ADM itself---even understood as a process---operates not on its own, but as part of a wider socio-technical system.

Breaking down this ADM process into its technical and organisational
elements allows us to systematically consider how contextually appropriate record-keeping, logging, and other documentary mechanisms at each stage of the process can allow for the process as a whole to be reviewed. This assists with understanding how and why systems and processes are functioning and how particular decisions are made, offering opportunities to identify points of intervention and control across the ADM process (such as
interventions by organisations at the point of decision-making to manage various risks or external interventions by or at the behest of regulators and others). We focus on ADM involving ML, 
but the approach can potentially apply across algorithmic systems.

As such, a reviewability framework potentially offers a useful and holistic form of transparency to support meaningful accountability for ADM. 
And, while producing reviewable processes is already achievable, there are also numerous research opportunities in this space. 
We therefore integrate related research into a reviewability framework and highlight areas requiring further attention. 

In doing so, we make three key contributions: \textit{(i)} 
drawing from the approach to accountable decision-making of well-established legal frameworks to consider ADM as a broad socio-technical process; \textit{(ii)} taking a holistic view of that process from system conception through to consequences and investigation, and \textit{(iii)} providing a systematic framework for determining the 
technical and organisational record-keeping and logging mechanisms for that process, to provide contextually appropriate information supporting meaningful accountability for algorithmic systems. 

First (\S\ref{sec:adm}), we discuss approaches to ADM accountability that focus narrowly on models (\S\ref{sec:modellimitations}), arguing that ADM should instead be understood as a broader socio-technical process to support different forms of accountability (\S\ref{sec:accountableprocess}). Next (\S\ref{sec:reviewability}), we elaborate on our concept of reviewability as a means of achieving this, describing its origins in our work on applying administrative law to public sector ADM (\S\ref{sec:adminlaw}) and setting out the concept of reviewability, what it involves, its general applicability, and its potential benefits (\S\ref{sec:reviewableadm}). 
We then (\S\ref{sec:framework}) set out a systematic framework for ML-driven ADM  to assist in implementing technical and organisational record-keeping and logging mechanisms across the key stages of that process (\S\ref{sec:commissioning}-\ref{sec:investigation}). 
Finally (\S\ref{sec:implementation}), we discuss some challenges for implementation and future directions for research. In all, we argue, reviewability can provide a practical, systematic, legally-grounded approach to producing useful transparency, through the recording of contextually appropriate information that supports meaningful accountability. 
 \section{Accountable Algorithmic Systems}
\label{sec:adm}

We use \textit{automated and algorithmic decision-making} (ADM) to mean decisions about natural or legal persons, their rights, interests, or entitlements made by other natural or legal persons using automated processes. These automated processes
can either directly produce a decision or produce information on which a human decision-maker subsequently bases their decision in whole or in part. {They may also directly entail some subsequent action or process, perhaps connecting to other automated systems~\cite{singh2019}.} These outcomes may or may not have legal or similarly significant effects for the subject of the decision. Our use contrasts with the EU's General Data Protection Regulation (`GDPR'), which applies higher standards 
only to \textit{solely} automated decision-making which does produces such effects~\cite{gdpr2016}. 

Though discussions of ADM often focus on technical concerns, ADM in reality involves complex \textit{algorithmic systems} -- socio-technical "arrangements of people and code"~\cite{seaver2013}. Technical systems are only one part of a broader assemblage of human and technical elements. Many algorithmic systems for decision-making may involve machine learning (ML) models to produce decisions, produce information on which decisions are subsequently made, or manage other aspects that themselves make decisions. We focus on ADM where ML models produce either decisions or information upon which decisions are made.

Increased use of ADM in both public and private sectors has led to concerns about biases, errors, malfunctions, profiling, data protection issues, and changing power relations, amongst others~\cite{pAI2020}. For instance, regulators investigated Apple Card after complaints that the algorithm was assigning higher credit limits to men than to women~\cite{vigdor_2019}. Amazon's recruiting algorithm was reportedly shut down because it was discriminating against women~\cite{dastin_2018}. In Australia, an applicant was denied access to information on how automated rental subsidies were calculated, as the system was third-party and protected intellectual property~\cite{adm_subsidy_case_2020}.

As a result, calls have grown for ADM to better support various forms of investigation, assessment, and redress. However, much research on improving the accountability of ADM has focused on the models that drive decisions, rather than on broader technical and organisational processes of which models are only one part.

\subsection{Accountability in ADM}

Our understanding of accountability in ADM aligns with Bovens’s work on accountability~\cite{bovens2006}, as interpreted for algorithmic contexts by Wieringa~\cite{wieringa2020account}. In Bovens’s model, accountability involves an \textit{actor} from whom an account is given, a \textit{forum} to whom it is given, and a \textit{relationship of accountability} between them.  Also considered is the \textit{nature of the account} itself, and any \textit{consequences} flowing from that account. For accountability to be meaningful, information given by the actor should support \textit{effective} deliberation and discussion by the forum and the imposition of any  consequences by the forum on the actor~\cite{wieringa2020account} (such as legal remedies or interventions to correct process malfunction). 

Although accountability can be thought of abstractly in general terms, in practice it is highly contextual -- different actors will likely be accountable for different aspects of the ADM process depending on what is to be accounted for, and the kinds, levels, and formats of information needed for a relevant and appropriate account will depend heavily on the forum to whom it is owed~\cite{bovens2006,wieringa2020account}. Bovens suggests that there are at least five kinds of accountability relationship, depending on the actor and the forum~\cite{bovens2006}: \textit{political} accountability (to elected representatives and so on), \textit{legal} accountability (to courts), \textit{administrative} accountability (to auditors and regulators), \textit{professional} accountability (to internal and external peers), and \textit{social} accountability (to civil society and individuals). 

\subsubsection*{Opacity in algorithmic systems.}
\label{sec:algopacity}

Achieving meaningful accountability of ADM is difficult. Commercial considerations and complex decision-making process involving ML often entail considerable opacity. 
Burrell identifies three forms of algorithmic opacity~\cite{burrell2016}; while Burrell refers primarily to model opacity, these also relate to broader ADM processes. That is,  the details of proprietary datasets, models, systems, and processes can be deliberately concealed to protect commercial interests (what Danaher calls ‘\textit{intentional opacity}’~\cite{danaher2016}). Details of ADM processes may be incomprehensible without relevant technical knowledge (‘\textit{illiterate opacity}’~\cite{danaher2016}). And the mismatch between the complex, mathematical nature of ML and human forms of reasoning makes models themselves difficult for even the technically literate to understand (‘\textit{intrinsic opacity}’~\cite{danaher2016}). Multiple forms of opacity can combine – it may be that data, models, systems, and processes are concealed for intellectual property reasons (e.g.~\cite{adm_subsidy_case_2020}), and that, even if they were not, the models would be incomprehensible. Table~\ref{tab:opacit} presents some of these types of opacity.

ML systems have sometimes been described as a `black box', which {might be} a choice made to deliberately obscure~\cite{foryciarz2020} (a form of intentional opacity). However, understanding ADM as a socio-technical process, we argue that there is another form of opacity -- \textit{unwitting} opacity, where those responsible for designing, developing, deploying, and using systems simply don't think to record relevant organisational aspects
of ADM processes (perhaps unaware of their relevance for meaningful accountability~\cite{anonVRST}). By implementing technical and organisational record-keeping and logging mechanisms that allow  processes to be holistically interrogated, intentional and unwitting opacity could be substantially addressed. 

However, there are also risks in providing \textit{too much} information. As well as the forms of opacity discussed, Stohl et al identify two further forms stemming from the ‘transparency paradox’~\cite{stohl2016}, where too much visibility actually reduces transparency. That is, opacity can result from providing too much or the wrong kind of information, or information presented inaccessibly, whether deliberately to obscure (‘\textit{strategic opacity}’) or because the forum's needs haven’t been considered (‘\textit{inadvertent opacity}’)~\cite{stohl2016}.
To avoid these forms of opacity, accountable ADM should provide not just any information about the technical and organisational elements of the process, but the right kind of information about aspects of the process that are relevant to the possible accountability relationships, presented in the appropriate way for the likely forums.

\begin{table}[]
    \scriptsize
    \begin{tabular}{| p{1.75cm} | p{6.2cm}|}
    \hline
    \textbf{Type of opacity}     & \textbf{Cause}\\
    \hline
    Intentional opacity & Details of processes concealed for commercial reasons \cite{burrell2016,danaher2016}\\
    Illiterate opacity  & Processes incomprehensible without technological literacy \cite{burrell2016,danaher2016}\\
    Intrinsic opacity   & Incompatibility between machine and human reasoning \cite{burrell2016,danaher2016}\\
    Unwitting opacity   & Unawareness of relevance of broader processes for accountability~[\S\ref{sec:algopacity}]\\
    Strategic opacity   & Process information deliberately presented in an inaccessible way~\cite{stohl2016}\\
    Inadvertent opacity & Information unintentionally presented in an inaccessible way  \cite{stohl2016}\\  \hline
    \end{tabular}
    \caption{Some types of opacity in algorithmic systems. }
    \label{tab:opacit}
\end{table}

\subsection{Limitations of model-focused mechanisms}
\label{sec:modellimitations}

However, considerable technical and legal research has focused on mechanisms to address forms of illiterate and intrinsic opacity by making models more transparent or understandable in some way, rather than looking more holistically across ADM processes. In focusing on model-centric mechanisms,  forms of opacity relating to other aspects of the process, potentially more 
relevant for accountability, have been under-discussed. This emphasis on models may itself even contribute to unwitting opacity by obscuring the need for mechanisms to address other aspects of those processes.

Proposals typically focus either on assisting those responsible for models to understand their functioning or on providing information to support other oversight mechanisms. This includes proposals for transparency of algorithms themselves, making code, model benchmarks, or other aspects of the model lifecycle available for scrutiny~\cite{pasquale2015, commons2018, diakopolous2016, burrell2016, mitchell2019model,partnershipai}. Some  have argued for equipping the general public with skills and knowledge to understand how models work~\cite{burrell2016}, or  that journalists could reverse engineer algorithms to inform the public about their workings~\cite{diakopolous2013, diakopolous2015}.

An increasing focus of research in recent years has been human interpretable explanations of the workings of ML models. Some proposals have seemingly been prompted by academic debate about the existence, nature, extent, and utility of a so-called ‘right to an explanation’~\cite{goodman2016, wachter2017, selbst2017, malgieri2017, edwards2017} in GDPR, which was passed in 2016. Surveys of the research landscape show a vast number of mechanisms for developing `interpretable' or `explainable' ML systems have been proposed, seeking to provide explanations and other interpretable accounts of model behaviour to a range of forums, many of which have come in the last few years~\cite{guidotti2018, arrieta2020}.

While such proposals and mechanisms have their place, many proposals---whether intended to inform the general public, system developers, regulators, or others---address models themselves, rather than  broader decision-making processes \textit{of which models are but one element}. Though explanations can assist some specific concerns, like model engineering~\cite{bhatt2020}, explanations or other approaches focused on how the model itself works or has reached a particular outcome may miss much of what is important ~\cite{wieringa2020account}. 

\subsubsection{Meaningful accountability needs more than models.}

~As ADM is a socio-technical process, with both human and technical elements, ‘accountable ADM’ should not just involve making models themselves explainable or transparent in some way. Models are only one aspect of ADM and cannot by themselves provide accounts of the process as a whole~\cite{singh2019}, whether to technical, legal, or other forums. In practice, other aspects of model development and related technical elements require consideration~\cite{domingos2012,partnershipai,singh2019}. Moreover, technical elements of the process cannot provide sufficient information about its human aspects. Though explanations can give some information about model functioning, problems with a process or a decision often originate outside of the model itself: in the purposes for which it is used, improper assumptions in design and use, and other human and organisational factors around the model. 

Moreover, particularly with explanations, model-focused approaches often unduly burden subjects of decisions with understanding and challenging them. Equipping people with skills to understand the models they encounter in their lives (as proposed by some) may seem attractive to technologists and those in technology policy, but much of the public are unlikely to have quite as much interest in ML's inner-workings. Rather, people are concerned with broader aspects of ADM: the purposes, roles, and outcomes of decision-making processes as a whole, as it is the whole process that affects them rather than simply certain technical aspects of it~\cite{edwards2017}. More generally, it is not entirely clear why the  public should \textit{have} to understand and evaluate the models that make decisions about them, particularly as they will often have little real choice but to subject themselves to those decisions. Moreover, by focusing on models and holding individuals themselves responsible for understanding how they work, broader technical and organisational processes and other systemic issues are obscured, and attention is diverted from other (perhaps more relevant) factors~\cite{ananny2016, edwards2017}.

While model-focused mechanisms are an important area of study, it is therefore important also to consider the broader processes of which models are only one part~\cite{wieringa2020account,singh2019}. As Ananny and Crawford argue, “rather than privileging a type of accountability that needs to look inside systems, … we [should] instead hold systems accountable by looking across them”~\cite{ananny2016}, recognising that these are socio-technical processes with both technical and human elements. Accountability also must consider algorithmic systems as they are situated -- not just technical features, but how they affect different people in their contexts of use~\cite{katell2020}. 

\subsection{Accountable ADM as a process}
\label{sec:accountableprocess}

Meaningful accountability thus requires a view of the whole socio-technical process~\cite{ananny2016, wieringa2020account, singh2019, lehr2017}, from  commissioning of the system; through design of the model, selection of training data, and training and testing procedures; to making individual decisions; and on to the effects of those decisions and any subsequent investigations. For Ananny and Crawford, as for us, accountability of these processes “requires not just seeing inside any one component of an assemblage but understanding how it works as a system”~\cite{ananny2016}. 

Understanding ADM as a process allows us to identify points across that process for review of relevant human and technical factors by the appropriate forum, and, if necessary, to intervene to correct, mitigate, or otherwise address (potential) problems. This depends on targeted record-keeping and logging mechanisms to provide useful transparency by capturing technical \textit{and} human (organisational) elements across the whole process -- not necessarily for those subject to decisions, but to facilitate understanding of the functioning of the algorithmic system as a whole and to enable meaningful accounts to and oversight more generally by designers, developers, deployers, and overseers. 

That accountable ML/ADM needs a view beyond technical to organisational elements has seen growing acceptance in recent years (e.g.,~\cite{wieringa2020account, ananny2016, eurocom2020, adalovelace2020, koene2019, ico2019, partnershipai, demos2020, raji2020, ico2020}). However, initiatives aimed at working towards this often lack systematic understandings either of ML as a broad process or of how to produce useful transparency to support meaningful accountability across that whole process. Instead, proposed approaches have often been piecemeal – identifying particular aspects of that process thought to be in some way problematic. While model-focused approaches support too narrow a form of accountability, piecemeal transparency of broader decision-making processes may not bring as much benefit as is hoped. Without a systematic understanding of the ADM process and how to achieve useful transparency across it, transparency mechanisms for specific aspects of that process won't necessarily give the holistic view required, so might not provide information relevant to a particular account. Moreover, incomplete records showing a problem could mislead, suggesting that issues lie in the wrong place. 

Conversely, while documentation is important~\cite{partnershipai}, there are also risks in mechanisms that provide \textit{too much} information by essentially recording and disclosing everything indiscriminately. Privacy and commercial or state surveillance concerns are raised by recording vast amounts of information on decision-making processes. And, again, providing too much information, or providing it in the wrong way, risks producing its own kinds of strategic or inadvertent opacity (the ‘transparency paradox’~\cite{stohl2016}). 

Moreover, as previously discussed, accountability is in practice highly contextual. Which kind of accountability relationship (political, legal, administrative, professional, or social) is relevant will influence the information the actor should provide. Information useful for regulators overseeing compliance with financial services regulations might differ from what someone overseeing their own systems needs to assess compliance with their internal policies, for instance. What subjects of decisions might find useful may differ from what courts need to assign liability for harm caused by decision-making processes. And how information is presented affects how well it can support meaningful accountability, and will again vary depending on the nature of the account and to whom it is owed~\cite{stohl2016}. Crucially, a meaningful account cannot be given unless the forum can understand and critically engage with the subject matter~\cite{wieringa2020account, kemper2018}.

\subsubsection*{Supporting accountability through contextually appropriate information.}

Following the above, we argue that ADM processes that support meaningful accountability require mechanisms for recording and providing \textbf{contextually appropriate} information so as to provide for \textit{useful} transparency. That is, accountable ADM is a process that requires transparency mechanisms targeted at particular aspects of that process to provide information which is:\\
\textit{(i)} \textbf{relevant} to the accountability relationships involved (i.e. from and to whom and in what form is an account likely to be owed);\\ 
\textit{(ii)} \textbf{accurate} in that it is correct, complete, and representative;\\ 
\textit{(iii)} \textbf{proportionate} to the level of transparency required (i.e. what granularity of information and degree of knowledge is likely to be needed about the operation of the process); and\\ \textit{(iv)} \textbf{comprehensible} by those to whom an account is likely to be owed (i.e. how to present information so as to be understandable).
A way of understanding transparency contextually and holistically across the ADM process so as to facilitate different forms of accountability as appropriate is therefore necessary. To achieve that, a more systematic approach to providing useful transparency that facilitates meaningful accountability is needed, as we now describe.
 \section{Revewability}
\label{sec:reviewability}

To support meaningful accountability, we argue that ADM processes should be designed and developed to be \textit{reviewable}. Reviewability as a general concept  involves technical and organisational record-keeping and logging mechanisms that expose the contextually appropriate information needed to assess algorithmic systems, their context, and their outputs for legal compliance, whether they are functioning within expected or desired parameters, or for any other form of assessment relevant to various accountability relationships.

\begin{figure}[h]
  \begin{center}
    \includegraphics[width=0.9\columnwidth]{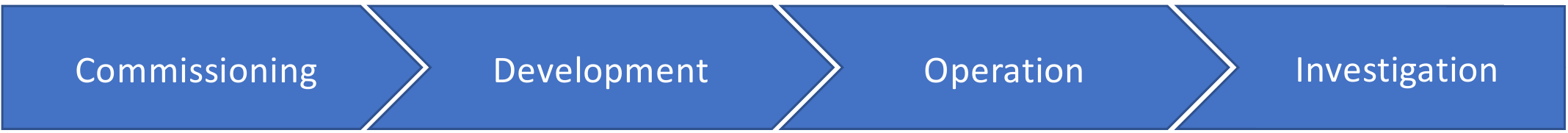}
  \end{center}
  \setlength{\belowcaptionskip}{-10pt} \caption{Conceptual view of the reviewability framework. \Small{Contextually appropriate technical and organisational information should be captured in all stages. \textit{Commissioning} refers to all that causes the overall system to be brought into existence - e.g. its nature and scope, legal basis, compliance assessment, procurement. \textit{Development} concerns details regarding system construction, encompassing the design and development of the technology and the relevant business processes. \textit{Operation} concerns details of use, including its application to particular scenarios (inputs), as well as deployment specifics, and details of system behaviour, business workflows, etc. \textit{Investigation} broadly concerns the that supporting investigation of the overall process(es), including evaluation metrics, interventions, remediations, etc. }
}
  \label{fig:revgen}
\end{figure}

In the context of ADM, reviewability seeks to provide a holistic view of the technical and organisational elements involved in producing an automated decision, considering factors both at design time \textit{and} at runtime. The commission, design, deployment, and use of ADM processes, as well as the consequences of use and auditing and investigation of those processes are all within scope of reviewability (see Fig.\ref{fig:revgen}). This approach is derived from English administrative law, which is concerned primarily with holding human decision-making in the public sector to account.

\subsection{Administrative law and ADM}
\label{sec:adminlaw}

Administrative law has developed to contend with the opacity and complexity of human decision-making, maintaining standards for even the most consequential decisions of life and death. With a few exceptions (around discrimination, for instance), administrative law as a framework and judicial review as a form oversight are generally not concerned with the merits of decisions themselves. 
Rather, they are concerned with the nature, quality, and legality of the decision-making \textit{process}, either in a general sense or as it relates to a particular decision. That is to say, courts are generally not concerned with whether a particular outcome is right, but whether the process that produced that outcome was correct. In administrative law, \textit{there is no general duty to give reasons (or explanations)}, but decision-makers must act in line with long-established principles of good administration throughout that process.

In administrative law, various aspects of the decision-making process are considered both discretely and together, allowing principles to apply to those aspects to---in theory---ensure good decision-making. For instance, nominated decision-makers cannot delegate decisions entirely to someone else, though can take advice into account. Decision-makers must consider all information relevant to a decision and cannot consider any irrelevant information; nor can they consider relevant information that is factually inaccurate. Decision-makers cannot give even the appearance of bias in making a decision. Decision-makers cannot unlawfully discriminate on a protected characteristic. Although not giving reasons for consequential decisions can sometimes be unlawful, reasons given---unless inadequate---are usually not themselves the basis for finding that a decision was made unlawfully (though they can give insight into how a decision-maker approached a decision). 
Judicial review instead assesses various aspects of the process for compliance with administrative law’s principles. 

Our previous work considered how administrative law and judicial review can apply to public sector ADM~\cite{anon2019}. In doing so, we drew on administrative law’s view of human decision-making as a process beginning before the decision and with consequences that resonate afterwards to show how these principles can apply to different aspects of ADM. Applying this understanding of decision-making as a process to ADM and identifying points at which the law’s principles are relevant potentially allows courts to review automated decisions made by public bodies without, for the most part, requiring explanations of the workings of the model itself. 

For example, administrative law's requirement that decision-makers consider all relevant information but no irrelevant information can apply to both model training and to making decisions using the model~\cite{anon2019}. This is particularly problematic for ADM given, the potential for proxies in ML \cite{prince2019proxy}, which can result in decisions based on factors that are not themselves directly relevant to the decision, and without considering factors that are in fact relevant~\cite{anon2019}. To assess compliance, reviewers do not need to understand the workings of any technical components -- they would instead consider the selection of factors for the model by its designer, the selection of training data, the data inputted for a decision, potentially inferences drawn by the model, and outputs produced by the model where they are subsequently relied upon by a human in making a decision. 

To reiterate, in this paper we do not seek to apply administrative law \textit{standards} to ADM in other sectors. Rather, we draw on and expand upon our previous work ---and administrative law's \textit{approach} to accountability of decision-making --- to develop a \textit{framework for} accountable ADM, applicable to accountability relationships of all kinds, in the public sector and elsewhere. This reviewability framework supports systematic and holistic assessment and review of ADM processes for compliance with \textit{any} technical, legal, ethical, and policy standards and requirements.

\subsection{Reviewable ADM}
\label{sec:reviewableadm}

Reviewability of ADM is concerned with exposing the whole decision-making process, potentially including: specifications and evaluations by those commissioning systems, decisions by engineers in developing systems; data used to train and test systems; training and testing procedures themselves; inferences drawn by the system while making automated decisions; and the fairness, effects, and lawfulness of those decisions in practice (see \S4). As such, reviewable ADM processes are those that systematically implement technical \textit{and} organisational record-keeping and logging mechanisms at all stages of commissioning, development, operation, and investigation to allow holistic review of the algorithmic system, its context, and outcomes. While ‘reviewability’ as a high-level concept has applications in various areas~\cite{anonIoT}, and is relevant for algorithmic systems \textit{in general}, it therefore takes an approach to transparency and accountability of ADM that goes beyond explanations or other mechanisms more narrowly focused on technical components.

Drawing from administrative law’s approach to identifying factors to assess at points in the human decision-making process, reviewability's systematic view of ADM focuses on points of review and intervention. As \S4 elaborates, these exist where people designing, developing, deploying, and using ADM take some action or decision relating to the process, or where technical components process data in some way. At these points, contextually appropriate information can be recorded about actions, decisions, or processing undertaken. This offers useful transparency of ADM processes both at design-time and---crucially---at run-time, providing information not just on how the algorithmic system was designed, developed, or intended to operate, but also on how it functions and what kinds of decisions its produces \textit{in practice}. By providing information at key points of the ADM process, reviewability assists in assessing individual decisions \textit{and} in determining whether the algorithmic system as a whole is functioning as intended or required.

This view across the whole process is important, given that accountability relationships and thus what it means to be accountable will differ depending on from and to whom an account is owed. Various actors at different stages across the same process may need to account to multiple forums – developers, regulators, courts, subjects of decisions, and so on. Implementing reviewability---with a systematic evaluation of what is contextually appropriate for various aspects of the decision-making process---helps ensure that relevant, accurate, proportionate, and comprehensible information is available to provide an account. By systematically implementing targeted technical and organisational record-keeping and logging mechanisms to enable the provision of contextually appropriate information about the process as a whole, reviewability thus supports meaningful accountability relationships between the multiple actors involved in the ADM process and various relevant forums.

 \section{A framework for reviewable ADM}
\label{sec:framework}

Reviewability offers a systematic approach to useful transparency by breaking down the ADM process into \textit{stages}---from conception of the system through to consequences and scrutiny---each consisting of a number of \textit{steps} (see Table~\ref{tab:modeladm}). These steps and stages can be considered discretely and together , underpinning a framework for developing and assessing reviewable ADM processes. At each stage there are opportunities to \textit{(i)} place limits on ADM and define (un)desirable behaviour or functionality, \textit{(ii)} implement contextually appropriate transparency mechanisms, \textit{(iii)} review compliance with those limits and general functioning, and \textit{(iv)} revise the process or take other action as required.

Reviewability thus supports a non-linear, iterative, and cyclical process of review, feedback, and revision, in line with the understanding of accountability discussed above and with our view of ADM as a process involving human and technical elements. 
Practitioners may move between steps non-linearly, depending on their role and the situation, as systems are developed, deployed, used, and revised. Not all steps will occur with each algorithmic system, and some will be more relevant than others, but at a high level these steps and stages will be common to many ADM processes.

Others have also discussed ML as a process involving multiple steps. Lehr and Ohm, for instance, propose two workflows with eight steps for helping the law understand ML: ``playing with the data'' (involving problem definition, data collection, data cleaning, summary statistics review, data partitioning, model selection, model training) and ``running model'' (involving model deployment) ~\cite{lehr2017}. Wieringa applies the Software Development Life Cycle model to divide the process into ‘ex ante’, ‘in media res’, and ‘ex post’ stages~\cite{wieringa2020account}. Suresh and Guttang split the model development lifecycle into six phases: data collection, preparation, model development, evaluation, post-processing, and deployment \cite{suresh2019framework}. 
The Partnership on AI also considers ML in stages, from ``system design and setup'' through ``maintenance'' and ``feedback'' once operational~\cite{partnershipai}. Guidance from the UK Information Commissioner's Office and the Alan Turing Institute group propose several tasks for producing explanations that relate to various aspects of the process~\cite{ico2020}.

These understandings of 
ML as a multi-step process are useful, but incomplete. First, while they acknowledge ML as socio-technical, they primarily focus on model-related issues, thereby not fully accounting for human and organisational aspects. 
Second, they miss some important aspects of ML processes prior to data collection and following deployment – procurement, impact assessments, consequences of decisions, and audit and assessment of that process (whether by those responsible or by an external overseer). Our framework, {grounded in administrative law's approach to accountable \textit{human} decision-making}, provides a more holistic view by adding two stages: ‘commissioning’  (involving procurement, problem definition, and impact assessment) and  ‘investigation’ (involving audit and disclosure), thereby encompassing more of the human decisions that are crucial to understanding the system.

\begin{table}[]
\footnotesize
\begin{tabular}{|p{3cm}|p{3cm}|}
\hline
\textbf{Stage}  & \textbf{Step}      \\
\hline
Commissioning   & Procurement        \\
                & Problem definition \\
                & Impact assessment  \\
                \hline
Model building  & Data collection    \\
                & Pre-processing     \\
                & Model training     \\
                & Model testing      \\
                \hline
Decision-making & Deployment         \\
                & Use                \\
                & Consequences       \\
                \hline
Investigation   & Audit              \\
                & Disclosure    \\
                \hline
\end{tabular}
\caption{Reviewability overview of a model-driven ADM process. 
  Each step entails record-keeping considerations. \label{tab:modeladm}}
\vspace{-0.5mm}
\end{table}

\subsubsection*{Reviewability in practice: a systematic approach.}

These stages and steps allow those responsible for ADM processes to consider accountability systematically. First, they should assess from which actor and to which forum accounts are likely to be owed at each. Forums, highly contextual, should be considered on a case-by-case basis when moving through the framework. 
Accountability of actors is likely to be dispersed, with obligations on multiple actors to explain or justify actions or decisions relating to aspects of the process for which they are responsible~\cite{wieringa2020account}. For identifying relevant actors, Wieringa’s work on applying Bovens’s model to algorithmic accountability can assist. In particular, Wieringa proposes three generic roles for actors in algorithmic systems~\cite{wieringa2020account} – \textit{decision-makers} (those responsible for deciding to use an algorithmic system and defining its specifications and other fundamental features), \textit{developers} (those responsible for specifics of developing the technical components to the required specification), and \textit{users} (those who use the system to produce a decision). We use \textit{managers} instead of \textit{decision-makers}, to avoid confusion with actors in the `decision-making' stage using the system to make decisions.

Having considered which actors and forums are relevant at each stage and step, those responsible for ADM processes should then consider which technical and organisational record-keeping and logging mechanisms could provide contextually appropriate information for those accountability relationships. 
We emphasise again that reviewability does not simply mean indiscriminate record-keeping at each step. 
Instead it is about targeted, useful transparency, providing information that is (i) relevant to the accountability relationships involved, (ii) accurate, complete, and representative, (iii) proportionate to the level of transparency likely to be required, and (iv) comprehensible by the relevant forums (\S\ref{sec:accountableprocess}). By considering each stage systematically in this way, the ADM process as a whole can be made reviewable, facilitating meaningful accounts of its operation to those to whom they may be owed. 

We now discuss what each stage involves, indicate which  actors are likely to be relevant at each stage and what kind of transparency mechanisms may be available or relevant at each step. Note we do not prescribe how to build reviewable systems, or to propose new technical or organisational solutions. Instead, we present a framework 
for systematically considering the information necessary for providing useful transparency, and thereby enabling meaningful accountability, in the context of a model-driven ADM process. We also indicate some factors that might warrant consideration.

\subsection{Commissioning}
\label{sec:commissioning}

Commissioning involves anything relevant to bringing the algorithmic system (and its constituent parts) into existence. 
At this stage, \textit{managers} are the actors likely to be particularly relevant.

\subsubsection{Problem definition.}
ADM exists for a purpose. Records relating to the aims of and rationale for the algorithmic system, giving insight into the values and norms behind its commissioning, development, and operation, are therefore relevant to various accountability relationships.
Documentation and other records of the system's aim, scope, and justification---what it will do, why it is required, and the role it will play---are worth considering.
Various impact assessment and procurement guidance documents reflect the need for clear specifications of these aspects (see below).
Also likely relevant is information regarding the decision-making processes or systems that ADM will subsume or replace. Information from business analysis or requirements engineering activities, common for many organisational technology undertakings~\cite{reqeng}, will often be pertinent, as will any documents such as minutes from board meetings, consultancy reports, and so on that involve discussions or decisions about the nature of the proposed system.

\subsubsection{Impact assessment.}
\label{sec:impactassess}

These involve assessing the potential implications and risks of an ADM system, and are important mechanisms for uncovering and mitigating potential problems. Assessments may encompass a range of concerns, such as legality and compliance~\cite{gdpr2016}, 
issues of discrimination and equality~\cite{canadaAlgAssess,pyper2020}, impacts on fundamental rights~\cite{janssen2020}, ethical issues~\cite{EuDenmark}, sustainability concerns~\cite{duboc2019}, amongst others. 
Some assessments are legally required -- GDPR, for instance, requires \textit{Data Protection Impact Assessments} (DPIAs) in various circumstances~\cite{gdpr2016}, including where there is an extensive evaluation of personal aspects that leads to decisions with significant effects (common for many ADM processes); other existing (e.g. \cite{canadaLaw})
and proposed (e.g. \cite{eurocom2020}) regulatory regimes also require assessments. Even where assessments are not legally required, they are good practice, and robust internal (in-organisation) 
assessment processes have value~\cite{raji2020}. 
There are many materials on various topics that can assist assessment undertakings~\cite{adalovelace2020,janssen2020,ainow}.

To facilitate accountability, records should be kept of any such assessment. These might include details of the actual assessment, and also the outcomes and any mitigation measures employed as a result. Other relevant information can include whether assessments were legally required and whether they led to an interaction with an oversight body (such as a regulator), as well as information on who conducted the assessment (an internal process or external organisation) and whether any advice was sought. Plans for ongoing monitoring and reassessment of the process are also likely to be relevant to various accountability relationships.

\subsubsection{Procurement.}
In practice, ADM will often involve some kind of procurement -- whether to obtain models or other technical components core to decision-making, data for training, technical components to support development or deployment, service arrangements for outsourcing business workflows, external consultancies for risk assessments, and so on. The nature of any procured product or service can influence the overall algorithmic system.

The importance of a robust procurement processes for algorithmic systems is well-recognised, with a range of guidance regarding this (e.g.~\cite{UKprocurementguideline,WEFprocurementguideline}). 
Records relating to procurement will likely serve various accountability relationships. This will often include details of contractual arrangements, tender documents, design specifications, quality assurance measures, and so on. Details of suppliers, including any due-diligence performed, may also be relevant. Any salient characteristics of what is being procured (e.g. test or acceptance criteria) are often best defined by the managers or documented as part of arrangements with the supplier. There should also be suitable mechanisms in place for suppliers to be audited or compelled to provide information as required (cf.~\cite{adm_subsidy_case_2020}).

Where procurement entails service engagements---e.g.  outsourcing of business processes, or engaging cloud  or other infrastructure services---details of the arrangement and terms of service will likely be important, as will levels-of-service guarantees, agreements for audit and inspection, etc. In future, third-party providers may adopt approaches such as service `factsheets'~\cite{arnold2019factsheets} that describe aspects of their service beyond technical capabilities.
Where procurement relates to technical `products', such as libraries, toolkits, test-kits, and software packages, details of the version, documentation, developing organisation, terms of use, availability, usage parameters or constraints, and licensing arrangements, may also aid review.

\subsection{Model building}
\label{sec:modelbuilding}

This stage involves the model development process, including related system design aspects; in particular, the human decisions in the model lifecycle. Any ADM process driven by models is in scope; if a system comprises several models, information on each model will be required, with further consideration of how the models interact with one another in practice systemically  as part of deployment. Much past work has focused on models and technical elements (\S\ref{sec:modellimitations}), and emerging regulatory frameworks 
explicitly reference record-keeping during model development~\cite{eurocom2020}. However, they often overlook the broader human dimensions of this stage.

In this stage prior to deployment, accountability of different actors is especially crucial. While \textit{developers} may be actively making decisions in the model build process, the implications of these decisions should be made clear, with appropriate oversight and documented approvals by \textit{managers}.

\begin{figure}[bth]
    \centering
    \includegraphics[width=.8\columnwidth]{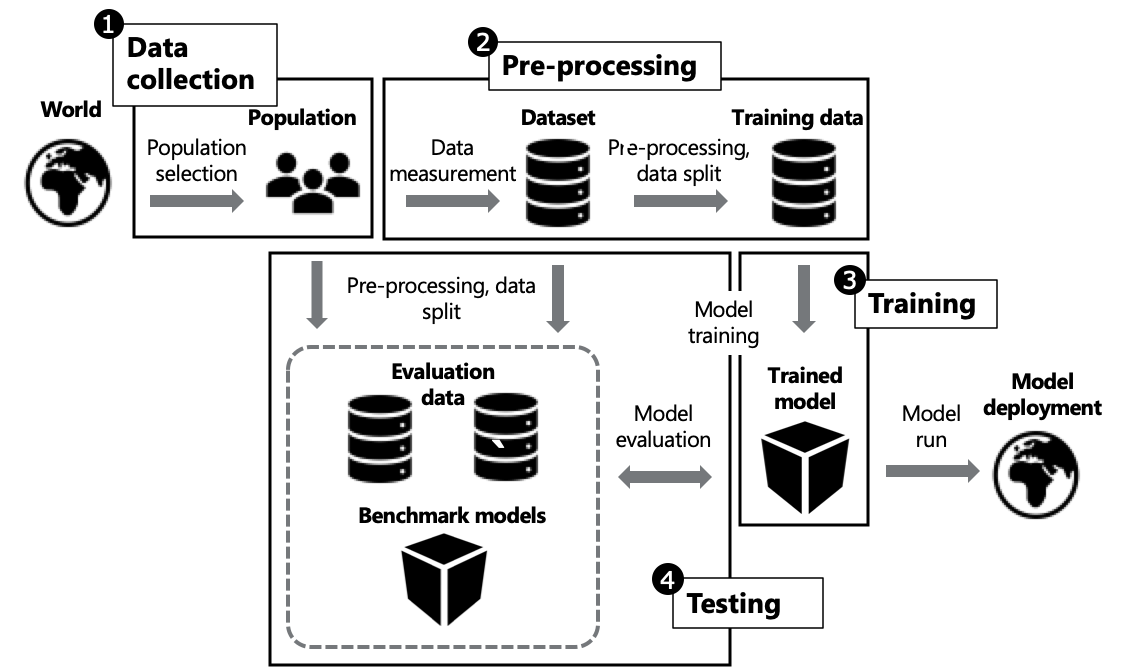}
    \setlength{\belowcaptionskip}{-5pt} \caption{Model build lifecycle: review process}
    \label{fig:lifecycle}
\end{figure}

Fig.~\ref{fig:lifecycle} demonstrates a typical ML model development lifecycle: 1) data collection, 2) pre-processing, 3) training, and 4) testing. At each lifecycle step, contextually appropriate records should be kept of qualitative and quantitative assessments of potential risk factors.

\subsubsection{Data collection.}

Before any model training, 
data must first be collected. This involves selecting a population from the world-at-large and relevant features with which to form a dataset. 
This can involve manual input of data, or surveys, or an automated process; such as through webpage scraping, or acquiring datasets. Often those undertaking data collection are separate from those responsible for the remaining model lifecycle; e.g. the collectors might be from a different institution or organisational unit.
Information on the decisions made as part of data collection is important for understanding the potential risks, limitations, and implications.
 
Details of the provenance of data, regarding its lineage and decisions made throughout its lifecycle---creation, collection, collation, processing, and sharing---is particularly important (both here and in the subsequent step)~\cite{singh2019}.
Gebru et al. propose standardised documentation processes for datasets~\cite{gebru2018datasheets}, especially useful for facilitating communication across those working with datasets. The~\textit{datasheet} contains questions for the data collector regarding: 1) motivation, 2) composition, 3) collection process, 4) pre-processing\slash cleaning\slash labelling, 5) uses, 6) distribution, and 7) maintenance. Questions 4-7 may not be fully answered by data collectors; while some processing may be done in collection (e.g. extracting text from a web site by removing markup), subsequent aspects would often be performed by the model developer with the specific use case in mind. Similarly, a ``data statement'' records relevant characteristics specific to text datasets for natural language processing ~\cite{bender2018data}.

The implications of how data is collected, how datasets are constructed, and how they relate to the system's potential risks are important and should be carefully considered. Additional questions may be needed to turn this information into actionable insights or to inform model design decisions. 
Contextually appropriate information relating to such assessments and questions will often need recording. Importantly, other information beyond that of the data itself---e.g. relating to other human decisions in data collection or to alternative options that were dismissed---are often also relevant. 

\subsubsection{Pre-processing.}

Pre-processing includes data cleansing (e.g. outlier detection, handling missing data or data inconsistencies), data wrangling\slash transformation, data merging, feature engineering\slash construction, data labelling, and feature selection~\cite{kotsiantis2006data,domingos2012}. Human decisions here could affect the model outcome; for instance, how missing values are imputed, changes to data structure or format, and selection of input features. Decisions in choosing and measuring features and labels can contribute to unintended measurement bias where they involve imperfect proxies for desired quantities~\cite{suresh2019framework}. For instance, using grade point average (GPA) to estimate student success is a decision that simplifies the latter with a proxy. As such, useful records would often relate to these various aspects of pre-processing, including (but not limited to) around which proxies were selected and why. 
Datasheets
include relevant questions to facilitate record-keeping for features in the dataset~\cite{gebru2018datasheets}; answers to questions about these aspects of model building should also consider the potential system-level implications of decisions. For example, the gap between the ``observed'' feature space and the ``decision'' space results in a mismeasurement of the target feature, especially in the presence of historical and structural discrimination; GPA, for instance, is an imperfect measure of success in high school~\cite{friedler2016possibility}. 

\subsubsection{Training.} 
~In preparation for model training, datasets are often split into training data, testing data, and (sometimes) validation data. In practice, some technical aspects of training and testing steps would be explicitly linked and iteratively performed, but there are aspects about the model training process that should be explicitly considered. Relevant information about the selection of training data (e.g. around ensuring that training data is representative of the dataset as a whole) and about the training processes should be recorded, as appropriate. Details about the workflow of model construction is often also important~\cite{singh2019}; this includes the machine learning approach(es) tried, tested (and perhaps, discarded), relevant aspects of tuning: hyperparameters, predictor variables, and coefficients (variable weights), and any other factors~\cite{schelter2017, partnershipai}. For example, model type and any parameters used in training, such as pruning methods in decision trees or choice of regularisation coefficient in lasso regressions, can be important to record, along with the results of the testing phase. These bear consideration as they provide the means not only to review the model and its process of creation, but also to enable some degree of model `reconstruction'.

\subsubsection{Testing.} 

Once a model is trained, it is tested to calculate relevant metrics. This may be iterative, as representation bias of testing data may be assessed against the population and the model may be retrained and re-tested to achieve the target metrics. Building on datasheets, `model cards for model reporting' offer standardised documentation procedures to communicate the performance characteristics of trained ML models~\cite{mitchell2019model}. The model card includes illustrative examples of questions in 9 categories: 1) Model Details, 2) Intended Use, 3) Factors, e.g. demographic or phenotypic groups, environmental conditions, technical attributes, 4) Metrics, 5) Evaluation Data, 6) Training Data, 7) Quantitative Analyses, e.g. of fairness, 8) Ethical Considerations, and 9) Caveats and Recommendations.

Model cards include some details on evaluation data and metrics~\cite{mitchell2019model}, including performance measures, but  models should be tested not only on accuracy but also safety, fairness, explainability, and security. Information relating to testing for these aspects is highly likely to be relevant to various accountability relationships. Any testing performed should be presented in the form of ``verifiable claims'' with audit trails and explanations of model predictions~\cite{brundage2020toward}. `FactSheets' have been proposed for ML models offered as a service, recording testing results to give confidence to the service users on the model~\cite{arnold2019factsheets}. Similar record-keeping mechanisms may be useful for model testing in other contexts. 

Selection of evaluation (or ``success'') metrics is relevant. Often these are subjective and value-laden.
In a well-known example, a US criminal recidivism model was accused of being racially biased~\cite{feller2016computer}, as, of defendants who ultimately did not reoffend, black people were more than twice as likely as white people to be classified as medium or high risk. The model's developers argued it was fair as, among defendants with the same risk score, the percentage of white and black defendants reoffended was broadly the same. 
Importantly, these two definitions of ``fairness'' are mathematically incompatible  -- in practice is it impossible to satisfy both simultaneously \cite{feller2016computer}. 
This illustrates the importance of recording how objectives, such as fairness, are defined, quantified and operationalised, so that they can be scrutinised and debated.

\subsection{Decision-making}
\label{sec:decisionmaking}

This stage concerns all operational aspects of ADM. This includes details of how the system is deployed and supported, and, importantly, all aspects leading up to a particular decision being made, and its consequences~\cite{singh2019}. Here, \textit{managers} and \textit{users} are most likely to be relevant, although  \textit{developers} may also be involved, particularly for the deployment step.

\subsubsection{Deployment.}

This concerns aspects of the process supporting the operation of the overall ADM system and its constituent parts. Information regarding deployment can help with understanding and verifying that appropriate mechanisms and procedures for supporting and maintaining the system are in place. Note that this step involves both a design and operational dimension; that is, building and testing the processes for deployment as well as supporting the ADM process `in production'.

From an organisational perspective, relevant information here relates to workflows and business processes relating to the algorithmic system. This encompasses, for instance, operating procedures, manuals, details of staff training and procedures relating to actually making decisions (using the system), as well as to operational support (maintaining the system). 
Also potentially relevant are records regarding provisioning and support of technical components, such as details  of the data and system pipelines~\cite{singh2019}, including model integration(s); storage, compute, and networking; scalability and security plans; logging mechanisms,  technical audit procedures, etc.

The management of the records and other data supporting useful transparency is another aspect bearing consideration. 
Details of the 
operating procedures, access management and integrity controls, encryption, and any other regimes in place to ensure that records and logs are appropriately managed may be relevant. 

\subsubsection{Use.}

This step concerns using the ADM system to actually make decisions {(whether the system `decides', or where a human makes the final decision)}. As such, it involves information on all aspects relevant to decisions actually being arrived at.

Records of a model's  inputs and outputs  are generally important, as are details of what and how information and feedback is presented to users. 
Parameters and metadata associated with a model's use also warrant consideration, as do operational records at a technical (systems log) level. Mechanisms for model interpretability or explanation or that otherwise describe how a model operates (see \S\ref{sec:modellimitations}) may also be relevant here.

However, this step also involves more than specifics of the {model's part in the} decision. Where {decision-making involves human users of the system}, or is otherwise manual, 
proper documentation arrangements are often required to record what occurs. 
For instance, many ADM systems will entail manual data entry, or will have a `human in the loop' working with the system to produce decisions -- information regarding this will often be pertinent. For technical processes, logging mechanisms can capture the details of inputs, outputs, and data processing\slash computation~\cite{singh2019}. Generally, `metadata'---including relevant time stamps, system or process versioning, records of any exceptional occurrences and operations, and so on--is also useful, often revealing potential issues. 

Note that for effective reviewability, it is essential that this step involves recording \textit{what actually occurs}. This is because it will generally be insufficient to rely on what is supposed to happen -- i.e. according to the the pre-defined workflows, business practices, and systems specification\slash documentation -- given the propensity for these to differ from what actually occurs day-to-day.

\subsubsection{Consequences.}
Once a decision is made, it is generally important to have information of the subsequent and follow-on steps, such as quality assurance processes (and their outcomes) for a decision, as well  mechanisms that review sets of historical decisions. This might include checking for model skew, or any inappropriate discrimination or other behaviour that may be manifesting~\cite{singh2016}.
Moreover, once a decision is made (and `finalised'),  generally relevant are records and logs about any actions taken to give effect to the decision. This includes, for instance, details about how the decision is communicated, and the triggering of any new workflows (e.g. initiating the loan process, assigning a mortgage, and so forth).

Note that cascading consequences can flow from a decision. Records of these are therefore important for better understanding the broader impacts and flow-on effects of decisions (see~\cite{singh2019}).

\subsection{Investigation}
\label{sec:investigation}

The investigation stage includes any {oversight or} investigatory activity, either internal {(e.g. by compliance teams)} or external (e.g., regulators and oversight agencies, civil society groups, individuals, and so on). 
\textit{Managers} are particularly relevant here, though \textit{developers} and \textit{users} may also play a role.

\subsubsection{Audit.}
Auditing processes are important to ensure that the entire ADM process works as intended. Audits might be conducted in-house, to evaluate and check that  decision-making processes are apt, {that procedures for human elements of the process are suitable and have been followed}, and that technical systems components are functioning correctly~\cite{ico2019,adalovelace2020}. Audits may also be external, whether by regulators checking compliance, investigators unpacking what led to a particular outcome, and so forth~\cite{adalovelace2020}. An entity may also seek to audit third-parties, such as organisations from whom they procure services, to ensure that arrangements and agreements are being met. 
Records of auditing activity facilitate scrutiny, and might include details of the audit, the basis or other reasons why it was undertaken, how it is conducted, and any findings. 
Where an audit leads to any subsequent actions or remedial response (e.g. system debugging, penalty actions), details of these should also be recorded. 
Given the potential sensitivity of audit data, records should be kept regarding how, when, why and by whom it was accessed.

\subsubsection{Disclosure.}
Disclosures make information about the ADM process available to others, and are therefore a key aspect of meaningful accountability to external forums. 
Organisations should have processes for making relevant records and logs available when requested and as contextually appropriate for the accountability relationships involved. 
This will generally require the collation and aggregation of information recorded in previous stages, ensuring that it is presented appropriately to support critical understanding by the forum.
Note that, in practice, many organisations appear ill-prepared for meeting their disclosure obligations~\cite{singhsecurity2019,wong2018}

Records about disclosures themselves can also be relevant -- both of the processes for disclosure as well as what was actually released, how  information was compiled, how it was delivered, in what format, to whom, and when. The basis for disclosure is also relevant -- was it legally required, e.g. as a result of an data access right under the GDPR~\cite{gdpr2016}, an order by a regulator, part of a legal proceeding, in line with established best practices, or even simply as an organisational choice. 
And, again, means for obtaining information from (third-party) suppliers may also require consideration.  

\section{Practicalities and future research}
\label{sec:implementation}

When implementing reviewability, we emphasise the importance of considering which mechanisms might provide useful transparency given the likely accountability relationships; i.e. (i) the actors from and the forums to whom an account for each stage of the process will likely be owed, and (ii) what kind of information relating to the process might be contextually appropriate for those relationships. Simply recording everything is not only often impractical---with \textit{vast} storage requirements~\cite{anonVRST, provstorage}---but could also result in a form of inadvertent opacity, where those assessing ADM processes (whether internally or externally) have too much information to sift through to be able to use it effectively (i.e. the ‘transparency paradox’~\cite{stohl2016}). Similarly, simply attempting to record all aspects without consideration may itself lead to transparency gaps, by virtue of the records appearing unrepresentative, unwittingly failing to capture particular aspects, and so forth.

Moreover, transparency can be harmful~\cite{ananny2016}. Recording information across a whole ADM process could bring substantial data protection and privacy risks, particularly if decisions concern people (especially where they are marginalised or vulnerable). Information recorded to facilitate review could potentially be personally revealing if wrongly disclosed and may be of interest to law enforcement agencies or surveillance programmes. To minimise the risks of harmful privacy breaches and surveillance, the relevance and proportionality aspects of assessing contextually appropriate information are crucial considerations.

Given these risks, good record management measures are essential. In practice, this means strong security regimes, potentially including technical measures (e.g. access controls, encryption, etc.) and organisational processes, to suitably manage and protect the information recorded. Details of the controls in place, as well as accesses and operations over this data, should also be recorded. There are practical challenges and opportunities for research in this space; including, for example, how to manage access to which data, in what circumstances, and for what purposes -- particularly given the range of stakeholders involved and reviewability's contextual nature. Similarly, given that this data relates to responsibilities, obligations, liabilities, etc, means for ensuring and verify the integrity and veracity of the  data are important. 

The presentation of information should also be considered. As previously discussed, too much information, or information presented without considering the forum's requirements, can produce inadvertent opacity~\cite{stohl2016}. Large amounts of data may be involved and raw records may require transformation to a form facilitating meaningful accountability. Mechanisms that assist this while being sensitive to the forum's needs are an area for further attention.

In this context, developing guidance, best practice, and standards for record-keeping and accountability more generally can assist. These can lead to more useful transparency regimes by raising the bar for organisational practices and increasing levels of consistency. There is scope for tools to assist organisations in identifying what is contextually appropriate, perhaps by indicating the needs of particular stakeholders, or particular sectorial considerations. There is also scope for specific guidance, e.g. around securing and managing records, appropriate record-keeping formats, standard forms of disclosure, and so forth. Technical  toolkits may also play a role in facilitating the logging and information to be extracted from technical components in a common manner~\cite{anonVRST}. Given that record-keeping requirements in a tech-context are increasingly backed by regulation~\cite{gdpr2016, eurocom2020}, there is a role for oversight bodies in helping define and shape what reviewability means `in practice'.

Finally, we emphasise that reviewability will not solve all problems with ADM. There are limits to transparency itself as a mechanism for supporting accountability~\cite{ananny2016,edwards2017,stohl2016}. Even reviewable systems may not capture broader business processes, the wider context of ADM processes, or the situations in which they are embedded. They may not capture interactions with other socio-technical processes, which form part of wider, complex, interconnected systems (although other mechanisms, such as \textit{decision provenance}~\cite{singh2019} could assist here). Nor will they provide information about business models or questions of power that come with automation of decision-making processes. It’s important also to remember that transparency has temporal limitations (i.e. “things change over time”~\cite{ananny2016}), and even contextually appropriate information is always open to subjective interpretation and contestation~\cite{ananny2016}. Reviewability must be understood as both a living and iterative process.

More fundamentally, even useful transparency may not provide a definitive understanding of how processes function, and will only go so far without substantive accountability and control mechanisms. 
Nevertheless, reviewability works to close a gap, by potentially providing information on (at least) the assumptions, decisions, and priorities of those responsible for designing, deploying, and using ADM and their practical effects.
As a mechanism for producing useful transparency across the ADM process, reviewability can, 
we argue, help support other (technical, legal, regulatory) accountability mechanisms that 
can
make a significant difference.
 \section{Conclusion}
\label{sec:conclusion}
Given ADM's rising prevalence, and risks of potential harm, methods for facilitating review and oversight are needed. Reviewability offers a legally-grounded, holistic, systematic, and practical framework for making algorithmic systems meaningfully accountable.
Through targeted technical and organisational record-keeping and logging mechanisms, reviewable ADM processes provide contextually appropriate information to support review and assessment both of individual decisions and of the process as a whole.

Reviewability cannot on its own address all potential problems with ADM; for that, a wider examination of its  political economy and socio-economic context is also needed, as are legal protections for individuals and groups. However, reviewability provides a way to gain a better understanding of ADM processes, not only for those employing ADM, but also for oversight bodies and those affected by decisions. Reviewable ADM could therefore potentially be better assessed for legal compliance and for decision quality, and can support those addressing more structural factors.

Research is clearly needed on implementation. The specifics of what record-keeping and logging might be appropriate at each step of the process, of what kind of information would be useful to retain, and of how this information can best be presented to forums to facilitate effective review of the algorithmic system’s operation depends on the system in question, the domain, and its purpose. However, as we have shown, existing mechanisms can be integrated into the reviewability framework, and we have indicated directions for future work to fill gaps. Although there is work to do, reviewability provides a practical way of developing meaningfully accountable ADM processes that can be implemented now.

\begin{acks}
The Compliant \& Accountable Systems Group acknowledges the financial support of the UK Engineering \& Physical Sciences Research Council (EP/P024394/1, EP/R033501/1), Aviva and Microsoft through the Microsoft Cloud Computing Research Centre.
\end{acks}

\bibliographystyle{ACM-Reference-Format}
\bibliography{sample-base}

\end{document}